\documentclass{sig-alternate}
\usepackage{booktabs}
\usepackage{threeparttable}
\usepackage{colortbl}
\usepackage[table]{xcolor}
\usepackage{array}
\usepackage[hyphens]{url}      

\begin{document}
%

\title{Privacy4ICTD in India: Exploring Perceptions, Attitudes and Awareness about ICT Use }
%
%
%
%
%

\numberofauthors{2} 
%
\author{
%
%
\alignauthor
Ponnurangam Kumaraguru\\
      \affaddr{IIIT - Delhi}\\
       \email{pk@iiitd.ac.in}
\alignauthor
Niharika Sachdeva\\
       \affaddr{IIIT - Delhi}\\
       \email{niharikas@iiitd.ac.in}
}

\maketitle
\begin{abstract}
Several ICT studies give anecdotal evidences showing privacy to be an area of concern that can influence adoption of technology in the developing world. However, in-depth understanding of end users' privacy attitudes and awareness is largely unexplored in developing countries such as India. We conducted a survey with 10,427 Indian citizens to bring forth various insights on privacy expectations and perceptions of end users. Our study explores end-users' privacy expectations on three ICT platforms --  mobile phones, OSN (Online Social Network), and government projects. Our results, though preliminary, show that users disproportionately consider financial details as personal information in comparison to medical records. Users heavily use mobile phones to store personal details and show high trust in mobile service providers for protecting the private data. However, users show concerns that mobile service provider may allow improper access of their personal information to third parties and government. We also find that female participants in the study were marginally more conscious of their privacy than males. To the best of our knowledge, this work presents the largest privacy study which benchmarks privacy perceptions among Indian citizens. Understanding users' privacy perceptions can help improve technology adoption and develop policies and laws for improving technology experience and enabling development for a better life in India.
\end{abstract}




\section{Introduction}

In recent years, Information and Communication Technologies (ICTs) have experienced increased interest of the research and design communities in creating technologies to sustain development. These technologies are used for various activities such as financial transaction, health services, entrepreneurship, governance, and entertainment~\cite{Dodson:2013:MGC:2516604.2516626, Kolko:2007:CIC:1242572.1242689, doi:10.1080/00220388.2011.621944, export:122814, Srinivasan:2013:CMR:2516604.2516625}. This use results in private information exchange and storage through technologies. Thus apart from enormous benefits, ICTs have raised privacy concerns among end-users in developing nations. For instance, research by Nokia shows a paradoxical view among young Indian users who on one hand love music, gaming and social networking, but also seek privacy for their online and mobile presence~\cite{Nokiasurvey}. Another article from The New York Times reports that people in Mumbai, India expressed a strong need for general privacy from neighbors and family~\cite{privacyhardtocomeby}. 

Privacy has been well studied in the developed countries~\cite{Brandeis:1890fk, lorriebeyondconcernatt, kumaraguru:privacy-indexes:-a-survey:2005:lrfkq}. Findings of the earlier studies bring forth various privacy concerns of end-user on the Internet use~\cite{Culnan_1999, Culnan2001}, profiling / tracking through Internet services~\cite{lorriebeyondconcernatt}, data collection and sharing by corporates~\cite{bellmaninternationaldifferences:2004}, surveillance by government~\cite{surveillanceusa}. In the developing world, the privacy understanding around ICTs is limited to anecdotal evidences provided by technology usability experiments and security evaluation performed by ICTD researchers~\cite{Ben-David:2011:CSD:1999927.1999939, Dodson:2013:MGC:2516604.2516626, Kolko:2007:CIC:1242572.1242689, Srinivasan:2013:CMR:2516604.2516625}. These studies show that technology poses novel privacy threats which may slow down or impede technology adoption~\cite{Ben-David:2011:CSD:1999927.1999939}. For instance, women are concerned about their privacy while taking assistance from expert users for using mobile phones~\cite{Dodson:2013:MGC:2516604.2516626}. 

Although, we do not argue that privacy is a bigger concern in developing nations than anywhere else, but we believe that the diverse landscape of technology use results in unique privacy issues in developing nations~\cite{kumaraguru:privacy-perceptions-in-in:2005:lrfkq}. Privacy concepts from the developed world may not be applicable to the developing world due to unique situations in developing regions. These situations include users' economic status, literacy, identity concepts, community sharing of resources, and lack of privacy laws ~\cite{Ben-David:2011:CSD:1999927.1999939}. Privacy concerns in developing regions like India are under-represented in the literature. To best of our knowledge, our study is the first empirical study conducted in India that provides insights on users' privacy attitudes for widely trusted ICT platforms. Through a survey of 10,427 respondents, conducted all across India, we present end-users' privacy attitudes and concerns about service providers and governments while using ICTs.

In this paper, we study three aspects of ICT services / platforms -- mobile phones, OSN (Online Social Networks such as Facebook), and government projects. These ICTs also show deep penetration in Indian society, in addition to increased interest of research and design community. There were 948 million number of wireless subscribers in India at the end of July, 2014~\cite{:2012fk}. India is also expected to have world's largest Facebook population in 2016 with a penetration rate of 7.7\% per year~\cite{Indiafacebookfact}. Currently, it has 92 million Facebook users (7.73\% of total Facebook users) who are spread across both major and small cities~\cite{Nayak:2014oq}. The e-governance facilities and projects such as Unique Identification (UID) reach throughout the country. Along with the benefits, services like UID, generated privacy concerns for both expert and novice users~\cite{Srinivasan:2013:CMR:2516604.2516625}. 


Privacy concerns in India result in many open questions and research gaps for ICT research. For instance, \emph{what users think is private information?}, \emph{with whom would they like to share the private information?} and \emph{are there any privacy concerns on how service providers handle private data of the users?} For such research questions, little is known about how the dynamics of population such as gender, influence the privacy expectation. In our work, we present exploratory analysis of respondents' perception to answer these questions. Privacy studies so far have mostly focused on the developed world~\cite{bellmaninternationaldifferences:2004, boniculturalaspects:2002, ecommerce1998, socialnetworkingpopularacrossglobe, Singh:2006fk}. Earlier studies also show that service providers and governments may be able to afford trained officials to maintain the privacy and protect data, however, end-users often lack the necessary resources to protect themselves from potential privacy threats~\cite{ion:home-is-safer-than-the-cl:2011:nrtys}. The research efforts in the developed world have helped to create privacy-preserving best practices and appropriate privacy policies that enable users to enjoy benefits of the technology without compromising their right to privacy~\cite{Culnan_1999, mcdonald2008cost}. We hope insights from our study will help ICT researchers and designers to include privacy needs of users in their designs while creating technology for developing nations such as India. Understanding privacy attitudes can play a crucial role in the acceptance of ICTs for development and to improve life in India. Some of our preliminary findings indicate that:
\begin{itemize}
\item{Participants \emph{disproportionately} consider financial details such as credit card numbers, annual house-hold income and passwords as PII (Personally Identifiable Information) in comparison to medical records.} 
\vspace{-2mm}
\item{Participants feel \emph{more comfortable} disclosing personal information such as financial details to governments than websites. They can disclose PII to mobile service providers even if not mandated.} 
\vspace{-2mm}
\item{Participants prefer sharing \emph{PII} such as annual household income and credit cards only with family members. However, they do not feel comfortable sharing passwords even with family.} 
\vspace{-2mm}
\item{Participants show \emph{little preparedness} to protect themselves from a privacy breach. Many participants accept friend request from unknown people such as people from their hometown and of the opposite gender.} 
\vspace{-6mm}
\item{We find statistically significant difference between men and women for \emph{what is PII}, and sharing PII with friends, family and society at large. Also, women are more conscious than men for accepting friend requests from unknown people.} 
\end{itemize}
\section{Related Work}
Many privacy surveys conducted in developed countries like the US, Europe, Australia, UK and Canada, show different privacy expectations of diverse stakeholders -- governments, corporates or marketers, and consumers~\cite{bellmaninternationaldifferences:2004, boniculturalaspects:2002, Dutton:2007kx, harrisinteractiveleadership, harrisinteractiveleadershipwave2, ecommerce1998, socialnetworkingpopularacrossglobe, Singh:2006fk}. These studies analyzed privacy concerns on various ICT platforms such as websites, OSN, and for services such as data outsourcing and marketing~\cite{Little:2011uq, parentsteensonlineprivacy, Canadianssurvey, socialnetworkingpopularacrossglobe, Stutzman:2012kx}. Surveys also analyzed awareness of data protection rights of the national protection authorities and users' perceptions on privacy of data transmission and legal framework~\cite{ Dara:2012ly, Survey-requested-by-Directorate-General-Justice-Freedom-and-Security-and-coordinated-by-Directorate-General-Communication:2008zr}. Westin designed some indices to classify people as ``fundamentalist,'' ``pragmatists'' and ``unconcerned'' denoting people of high, medium, and low privacy concerns respectively~\cite{kumaraguru:privacy-indexes:-a-survey:2005:lrfkq}. 
These studies provide a comprehensive understanding of user expectations and behavior in the developed countries, but the complex usage landscape of ICT in developing regions make it non-trivial to apply this knowledge in developing countries. 

ICT4D (Information and Communication Technology for Development) research studies various aspects of technology that influence users' behavior and expectation~\cite{Coceres:2012:ICT:2360739.2360772, export:122814}. These aspects include utility, usability, penetration, financial benefits, and effectiveness of ICTs to achieve the desired development goals~\cite{ doi:10.1080/00220388.2011.621944, Parikh:2007:DAD:1368559, Putnam:2010:SMI:2369220.2369252, export:122814}. These systems face a primary paradox that on the one side, personal information helps to improve interactions, aid information exchange, and enhance facilities. However, on the other side, the same information introduces risks, ranging from minor issues to extreme privacy threats. Several research works show anecdotal evidences that privacy is a potential issue influencing technology use, penetration, and effectiveness for use of ICT in development initiatives~\cite{Ben-David:2011:CSD:1999927.1999939, Dodson:2013:MGC:2516604.2516626, Kolko:2007:CIC:1242572.1242689, Srinivasan:2013:CMR:2516604.2516625}. For example, studies show privacy is a concern for communities that use mobile phone as a shared device and use different ICTs involving intermediate assistance from others~\cite{Dodson:2013:MGC:2516604.2516626}. For such technologies, intermediate assistance from experienced users created privacy issues for novice users. They found that privacy concerns go beyond content sharing and involve complex dynamics of social relations, power relations, and gender. 



Few studies analyze privacy perspective of participants who originate from developing regions such as India, but stay or study in foreign countries. These studies provide mixed evidences / insights about the Indian privacy perspective. For instance, one such study showed that compared with fellow American and Chinese, Indian student participants were least concerned about privacy on various OSN~\cite{reference8}. However, another study, which examined cross-national differences in the usage of OSNs between university students in India and the US, found that Indian American students showed behaviors significantly more individualist than the American students~\cite{Marshall:2008ys}. 

We found very few studies that analyzed the privacy perception of participants residing in India. These studies conducted interviews and surveys showing that Indian students had different views, and showed less privacy concerns than nationals of foreign origin like America and Switzerland~\cite{ion:home-is-safer-than-the-cl:2011:nrtys, kumaraguru:privacy-perceptions-in-in:2005:lrfkq}. Contrary to these observations, some studies showed that Indian inhabitants were more privacy conscious. For example, Indian knowledge workers expressed higher interpersonal privacy concerns than their US colleagues~\cite{Patil:2010ve}. The sample size from India in these studies was 407 and190 which is relatively small~\cite{ion:home-is-safer-than-the-cl:2011:nrtys, kumaraguru:privacy-perceptions-in-in:2005:lrfkq}. Thus, does not give an explicit depiction of privacy awareness and perceptions in India. We believe that an empirical analysis of end-users' attitudes and awareness can help to develop a clear understanding of privacy among Indian inhabitants. 

We found that little is known about users' understanding and expectation of privacy on ICT such as mobile phone and OSN. \emph{How users are concerned about their privacy} also remains largely unexplored in India. To fill the gap in understanding users' perceptions, we empirically explore users' beliefs about privacy expectation on ICT services and platforms such as mobile phones, OSN, and government initiatives such as UID. In particular, we study issues such as what personal information is stored on these platforms, and from whom privacy is needed. Understanding privacy expectations not only helps develop better policies, and laws but also helps develop better technologies. India is in the phase of adopting new policies, technologies and schemes such as UID. We hope that the privacy understanding from our survey will help develop such policies, technologies, and adopt this learning to ICT influence in India and similar regions. These outcomes can increase end-users' trust in technology and can help in mass adoption of ICTs that aim to improve life in developing countries. 

\section{Methodology}
In this section, we present our data collection methodology and data analysis techniques. We collected 10,427 survey responses from participants all around India. We were not required to go through an Institutional Review Board (IRB) approval process before conducting the study. However, the authors of this paper have previously been involved in studies with the US IRB approvals and have applied similar practices in this study. Participants were shown consent information if they agreed to respond in our survey or else they had a choice to not participate. 

 \subsection {Survey}
To understand users' privacy perceptions, we administered preliminary unstructured interviews and semi - structured Focused Group Discussions (FGD) with a geographically spread group of Indian users. These interviews and FGDs showed that users held privacy concerns primarily on three aspects - 1) Mobile phones usage, 2) OSN access, and 3) Government control on ICTs. Based on these inferences, we designed a survey to gain a better and an empirical understanding of users' privacy concern. The survey had 17 questions covering demographics, 16, 19, and 7 questions on mobile service providers, OSN services, and Government services respectively, and 26 Likert scale questions covering users' privacy attitudes. In the survey, we also included questions on general privacy expectations of the participants in the physical space. In total, the survey had 74 questions. We kept open choice option in various questions to allow users to specify any other preference that they will like to share with us. We did not collect any identifying information such as participant's name or address. However, we announced a raffle prize to incentivize participants to share their views with us. If participants chose to participate in this draw, they could share their contact details such as email id. Such a survey method approach has been used in earlier ICTD works to show the influence of technology~\cite{Kolko:2007:CIC:1242572.1242689}. 

To understand users' concern on ~personal ~information ~~stored~on mobile phones and OSN, we asked participants questions such as \emph{what is the personal information which you [users] don't mind storing in your mobile phone?} and \emph{what are the reasons for which you} [users] \emph{dont store personal information on your} [their] \emph{mobile phone?} To understand the concern on OSN, we asked questions such as \emph{if you} [user] \emph{receive a friendship request on your} [users'] \emph{most frequently used OSN, which of the following people, will you} [they] \emph{add as friends?} We also included questions regarding the privacy offered by mobile service providers and online social networking sites. 

\begin{figure}
\begin{center}
 \frame{\includegraphics[scale=0.20]{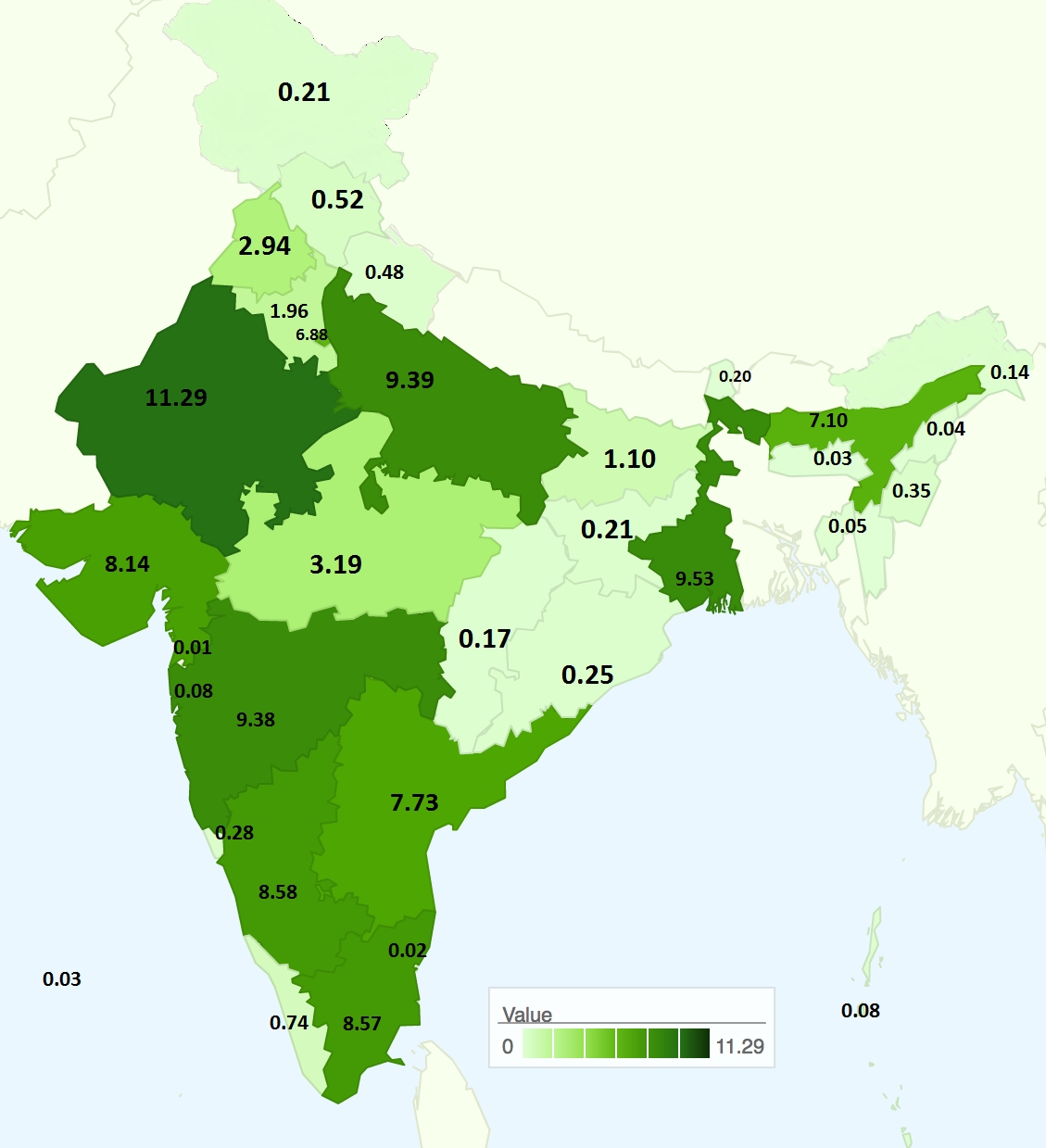}}
\caption{Percentage of sample from different states in India. Mizoram is the only with no representation in our sample.}
\label{fig:type}
\vspace{-9mm}
\end{center}
\end{figure}

The survey was developed in English as it is common business language used in urban and sub-urban regions of India. The technology penetration such as smartphones and OSN use (that enables sharing or storing of personal information) is high in these regions (urban and sub-urban). Therefore, users in these regions can experience diverse, demonstrative, and impactful privacy threats and concerns. 
Given the available resources, in total, we collected valid responses from 10,427 participants. Our sample consisted of respondents from different cities in India; we travelled to multiple cities for data collection and also took help from various organizations to collect data. Our survey consists of respondents from every state and union territory in the country, except for one state (Mizoram) (see Figure ~\ref{fig:type}). Table~\ref{tab:surveydemo} shows the demographics of the survey participants. We realize that number of males in our study are dominant, however, male and female ratio in our study, is representative ratio of both genders who use technology and are comfortable with English in the country. 

Before starting the data collection, we did a pilot study with 70 respondents. After minor changes, we collected data through different means such as web (Survey Monkey), and physical printouts. As our goal was to observe privacy expectations from different walks of life, we collected data from schools, universities, organizations, metro stations, households, malls, etc.  Majority (60\%) of the participants got to know about our survey through various market research organizations that we worked with to collect data and about 31\% of the respondents got to know it through ``friends and acquaintances." Majority of the participants (63\%) filled the survey in ``Public space e.g. shopping mall, cinema, market, and park" and about 28\% filled it in their campus or organization. 
 
Authors understand that this methodology is prone to self - selection bias and any other bias that may arise because of recruitment sources. Also, users who were comfortable using OSN, and / or mobile phones or government schemes like UID participated in our study. Therefore, authors do not make any claims on the representativeness of the dataset. Despite this, we believe that our data gives a fair representation of the users' expectation, and our findings can be reproduced for similar Indian population. Also, ours is the first such study, so no other representative data is available for comparisons. 

\begin{table}[!h]
\vspace{-3mm}
\centering
\small
\caption{\small{Demographics of participants, N $=$ 10,427. }\label{table:pi2013:demographics}}
\small
\begin{tabular}{p{5cm} p{1.55cm} r} 
\hline
\midrule
\raggedright \bf{Gender} & (\%)\\
\raggedright Male &67.57 \\
\raggedright Female &32.43\\
\midrule
\raggedright \bf{Age} (in years) &  (\%) \\
 \raggedright 18 to 24 &21.31\\
\raggedright  25 to 29 &32.20  \\
\raggedright  30 to 39  &25.90 \\
\raggedright  40 to 49 &14.09 \\
\raggedright  50 to 64 and above &4.96 \\
\midrule
\raggedright \bf{Education Completed} &  (\%)\\
 \raggedright Less than High school&9.75 \\
 \raggedright High school&24.74 \\
\raggedright  College Graduate& 46.16\\
\raggedright  Post Graduate &16.19\\
\raggedright  Doctorate & 1.34\\
\midrule
\raggedright \bf{Profession} & (\%)  \\
\raggedright Computer / IT related & 18.65\\
\raggedright Housewife & 13.25\\
\raggedright Manufacturing / Business & 28.77\\
\raggedright Professional (Doctor, CA, etc.) & 2.67\\ 
\raggedright Student& 20.20\\
\raggedright Teaching / Research& 4.75\\
\raggedright Other& 11.72\\
\hline
\hline
\end{tabular}
\vspace{-3mm}
\label{tab:surveydemo}
\end{table}


\subsection{Data Analysis}
We used various measures to determine if our participants had not given random responses to the survey. For this, we considered the time in which participants completed the survey and ignored surveys that were completed in unusually short duration. We asked same question with slightly different wording and checked if participants gave same responses. We removed surveys that showed inconsistent answers. After manually checking the answers, we were left with 10,427 responses from initial collection of 10,440 responses. We analyzed data using a series of statistical tests to check the linkage and relationship between various variables. Users could choose not to answer questions asking information that they did not feel comfortable sharing, therefore for such questions the number of respondents is less than 10,427. For analyzing dependent data, for e.g. evaluating response of the same participants for various variables, we used statistical test such as McNemar's test and Wilcoxon Signed-test  for Likert scale questions. We used Chi -- squared test wherever appropriate to show the difference between groups based on their age and gender. Appropriate corrections were applied where needed.  

\section{Results}
In this section, we show users' privacy attitudes, expectations and awareness. We discuss four themes: 1) Perceptions about personal information on mobile phones and OSN, 2)Trust in service operators and fear of surveillance, 3) Social sharing of private information, and 4) Privacy attitudes that can influence technology use.

\subsection{Personal information on mobile phones \& OSN}
Most ICT projects such as mobile ~finance and ~health ~~~projects require collection of users' personal information. Users' unwillingness to share this information can be an obstacle for such schemes. We now present: 1) Participants disproportionately consider passwords or financial detail as personal information that they will not like to share when compared with health details or physical details and 2) Participants heavily rely on mobile phone to store PII and share it on OSN.

Participants marked their response to the following question, \emph{ which of the following information is personal to you that you would NOT like to share?} Participants chose from a range of options such as annual household income, marital status, bank account details, mobile number, passwords, e-mail address, and physical details -- height, weight, eye color (See Table~\ref{tab:PIIwhat}). We found that disproportionately large number of participants perceived information such as password (88.39\%), and financial details such as credit card number (68.18\%), bank account details (64.63\%) as PII  that they will not like to share with others. Participants considered medical health records (27.17\%), and physical details (8.47\%) could be  shared with others. 

\begin{table}[!htbp]
\caption{\small{Participants' response to information that they consider as PII. All values are in percentage. N shows the number of participants who answered this question.}}
\small
\centering
\setlength{\extrarowheight}{2pt}
\begin{tabular}{p{6cm} r p{1.4cm} r}
\midrule
\textbf{N=10,377}&\textbf{(\%)}\\
\midrule
 \rowcolor {gray!16 }
\raggedright Annual house hold income &53.64\\
\raggedright Bank account details & 64.63\\
 \rowcolor {gray!16 }
\raggedright Credit card number & 68.18\\
\raggedright Health and medical history &27.17\\
 \rowcolor {gray!16 }
\raggedright Passport number& 64.45\\
\raggedright Passwords &88.39\\
 \rowcolor {gray!16 }
\raggedright Personal income &62.77\\
\raggedright Physical details - height, weight, eye colour & 8.47\\
\midrule
\hline
\end{tabular}
\label{tab:PIIwhat}
\end{table} 

Next, we analyzed how comfortable users felt sharing personal information on OSN. We asked participants what personal information they have shared on OSN. Participants said that they shared information such as photos (48.95\%) and videos (45.41\%) on OSN with friends (see Table ~\ref{tab:PIIOSN}). A significant number of participants marked sharing information such as location (45.95\%) and religious preference (40.20\%) with everyone on OSN. Furthermore, most participants reported use of mobile phone to store personal information such as videos and photographs (64.57\%). Few participants also agreed to use mobile phone to store personal information such as passwords (25\%), credit card number(s) / ATM card number(s) / PIN number(s) (26.2\%), date of birth and ID number (30.51\%). These percentages show participants used technology such as mobile phones and OSN to store personal information. However, they did not feel comfortable storing personal details such as passwords on mobile phones. Participants who did not save information on mobile phones said that they were concerned about somebody accessing the phone at work, or outdoors without permission (38.51\%) and somebody accessing the phone at home without permission (24.86\%). 

\begin{figure}[!htbp]
\begin{center}
\vspace{-5mm}
\includegraphics*[viewport= 2 2 800 300, scale=0.40]{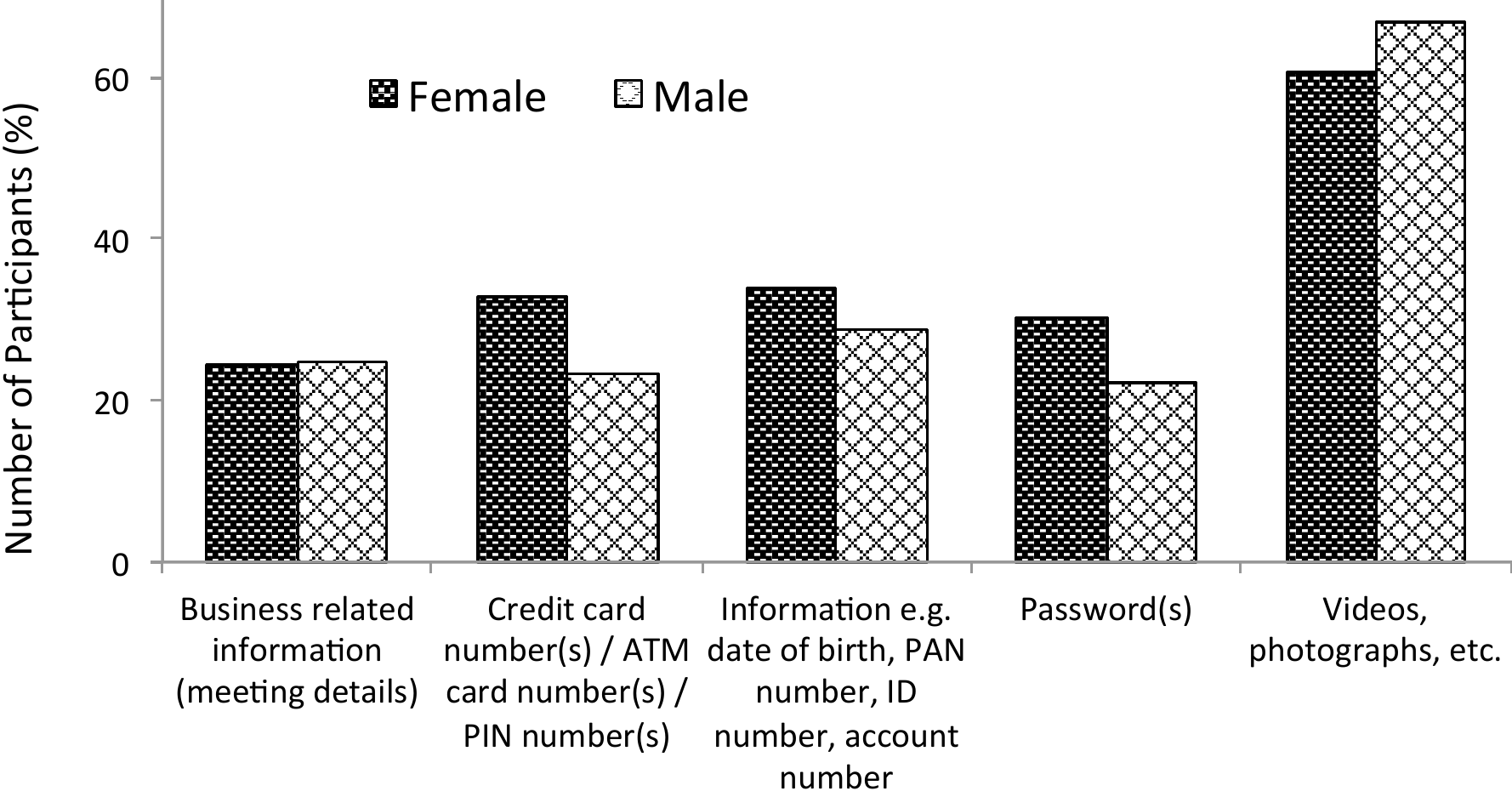}
\vspace{-3mm}
\caption{\small{Gender preference for storing PII on mobile phones. Females used mobile phones more than males to store personal information.}}
\label{fig:FM2}
\end{center}
\vspace{-5mm}
\end{figure}

We used Chi - squared test with holm corrections to examine if statistical difference exists between two genders on PII understanding and sharing of personal information. Our results show statistically significant difference between men and women on both a) Perception of \emph{what was PII} (Chi -sq test with holm corrections, p<0.01) and b) \emph{if they shared PII on OSN or stored PII on phone} (Chi -sq test with holm corrections, p<0.01). Figure~\ref{fig:FM2} shows gender preference for storing PII on mobile phones. Women used mobile phones more than men to store personal information. We found statistically significant difference between men  and women for saving information (shown in Figure~\ref{fig:FM2}) such as credit card number(s) / ATM card number(s) / PIN number(s) on mobile phone ($\chi^2$ = 55.723, df = 1, p-value < 0.0001). Similarly, we found statistical difference on gender's preference for storing passwords on mobile phones ($\chi^2$ = 40.4386, df = 1, p-value< 0.0001). 

\subsection{Trust in service operators and ~fear ~of ~~surveillance}
We analyzed participants' trust in the service operators for privacy, they provided. ~We ~found ~that ~participants ~showed: 1) Concern for personal data collected by service providers and governments and 2) High concern for improper access, i.e., disclosure of information by service providers to unauthorized parties~\cite{bellmaninternationaldifferences:2004}. 

\begin{table}[!htbp]
\vspace{-4mm}
\centering
\caption{\small{Participants' response for information that they have shared on OSN. All values are in percentage.}}
\small
\setlength{\extrarowheight}{2pt}
\begin{tabular}{ p{2.5cm} p{1.5cm}  p{1.0cm}  p{1.2cm} }
\midrule
\midrule          
& Not Shared & Friends & Everyone\\
\hline
      \rowcolor {gray!16 }
   Age   & 13.24 & 40.64  & 33.14  \\
    Date of Birth & 7.36  & 44.59 & 31.89  \\
        \rowcolor {gray!16 }
    E mail ID & 6.49  & 44.17 & 30.77  \\
    Gender & 2.97  & 32.49 & 45.95  \\
          \rowcolor {gray!16 }
    Location & 5.77  & 33.13  & 39.83\\
    Marital Status & 8.13  & 32.59 & 39.08  \\
         \rowcolor {gray!16 }

  \raggedright Other Profile Information e.g. education and work details & 5.95  & 40.34  & 30.37  \\

    Pictures / Photos & 7.18  & 48.95 & 21.02  \\
      \rowcolor {gray!16 }
    Religion & 8.96  & 32.18 & 40.20  \\
    Videos & 11.01 & 45.41 & 21.20 \\
    \midrule
    \hline
    \end{tabular}%
    \vspace{-4mm}
    \label{tab:PIIOSN}
\end{table}%

\subsubsection{Trust for data collection}
Participants marked how comfortable they felt sharing PII with websites and government. We found that participants felt more comfortable sharing financial information such as annual household income, bank account details, and credit card numbers with the government than with websites (See Table \ref{tab:PIIshared}). However, they felt more comfortable sharing information such as medical history, marital status, and mobile number with websites than the government.
We also asked participants if they would share personal information with mobile service providers if not mandatory (See Table \ref{table:ans4}). We found that participants felt comfortable sharing contact information such as address, ID proof / passport number, and contact number with mobile service providers, followed by websites. We also analyzed participants' confidence in websites and mobile service providers for maintaining privacy of the data they collected. About 21\% participants strongly agreed with the statement that websites can hinder privacy by collecting personal information. Percentage of participants (63.21\%) trusting mobile service providers was much higher than websites (See Table~\ref{tab:pi2013:govcop}).
\vspace{-4mm}
\begin{table}[!htbp]
\small
\centering
  \caption{\small{Participants shared various PII with websites and government. All values are in percentage.}} 
 \vspace{.2mm}
\begin{tabular}{p{4.0cm} p{1.3cm} p{1.6cm} }
 \hline
\midrule
	\raggedright  & \bf{Websites} &\bf{Government} \\
& (\%) &(\%)\\
\midrule
 \rowcolor {gray!16 }
\raggedright Annual household income &7.83&21.40\\
\raggedright Bank account details &3.02&14.14\\
 \rowcolor {gray!16 }
\raggedright Credit card number &3.08&5.05\\
\raggedright Date of birth & 32.96&12.85\\
 \rowcolor {gray!16 }
\raggedright Email address &33.38&9.98 \\
\raggedright Family details &20.98&11.57\\
 \rowcolor {gray!16 }
\raggedright Full name&42.33&10.09\\
\raggedright Health and medical history &16.64&5.97\\
 \rowcolor {gray!16 }
\raggedright Landline number &22.40&8.74\\
\raggedright Marital status&34.26&10.02\\
 \rowcolor {gray!16 }
\raggedright Mobile number &24.50&12.54\\
\raggedright Passport number&4.76&11.22\\
 \rowcolor {gray!15 }
\raggedright Passwords &2.79&1.01\\
\raggedright Personal income &4.20&6.16\\
 \rowcolor {gray!15 }
\raggedright Pictures and videos featuring self &21.67&3.47\\
\raggedright Physical details - height, weight, eye color & 26.15&6.14\\
 \rowcolor {gray!15 }
\raggedright Postal mailing address&30.73 &15.46\\
\raggedright Religion &39.87 &8.94\\
\midrule
\hline
\end{tabular}
\vspace{-3mm}
\label{tab:PIIshared}
\end{table}

\begin{table*}[!htbp]
 \centering
 \small
\caption{\small{Participants' response to data collection and handling by businesses on a Likert scale of 5. High trust on mobile service provider for data collection, whereas participants' showed lack of trust in websites. SA = Strongly Agree and A = Agree. $\mu$ represents mean rating on a likert scale of 5 and $\sigma$ represents the standard deviation. }}
\setlength{\extrarowheight}{2pt}
\begin{tabular}{p{4.8cm} p{1.8cm} p{1.4cm} p{2cm} p{1.5cm} p{1.2cm} p{1cm} p{1cm}}
\midrule
\midrule
\centering
\raggedright & \bf{Strongly agree/Agree} & \bf{Neutral} & \bf{Strongly Disagree/Disagree} & \bf{$\mu$ }&\bf{$\sigma$} & \bf{Male* (SA+A)} & \bf{Female* (SA+A)}\\
& &  & && & & \\
\midrule
\rowcolor {gray!15 } 
\raggedright Websites hinder privacy by collecting personal information (N=10,415) &71.22 &20.60 &7.25&2.16&0.85 &75.41&70.18\\
\raggedright Mobile service providers give reasonable protection to collected information (N=10,379) & 63.21 &20.86 &15.93 &2.42 & 0.97&58.98 & 65.62\\
\midrule
\midrule
\end{tabular}
\begin{tablenotes}[para,flushleft]
 \item{Note: All values are in percentages except mean and standard deviation. Both male and female participants show higher confidence in mobile service provider than websites. * shows p$>$ 0.05 (Mann Whitney Test)}
  \end{tablenotes}
  \vspace{-4mm}
\label{tab:pi2013:govcop}
\end{table*}


\begin{table}[!hb]
\vspace{-8mm}
 \centering
 \small
\caption{\small{Information shared with mobile service provider if not mandatory.}}
\setlength{\extrarowheight}{2pt}
\begin{tabular}{p{4cm}p{1.5cm}p{1cm}}
\midrule
\midrule
\bf{N=10,093}& \bf{Female} & \bf{Male}\\
&(\%)&(\%)\\
\midrule
 \rowcolor {gray!15 }
\raggedright Alternative address proof** & 39.8 & 43.99\\
\raggedright Another contact number & 33.82& 33.9\\
 \rowcolor {gray!15 }
\raggedright Educational qualification** & 31.11& 21.54\\
\raggedright ID proof** & 60.23& 70.44\\
 \rowcolor {gray!15 }
\raggedright Permanent address proof* & 32.24& 29.19\\
\raggedright Photograph(s)**& 54.69& 67.86\\
 \rowcolor {gray!15 }
\raggedright Proof of place of work & 17.85& 18.12\\
\raggedright Parents' details**& 10.55& 7.08\\
 \rowcolor {gray!15 }
\raggedright All of the above**& 12.09& 6.41\\
\raggedright None of the above*& 4.72& 3.7\\
 \rowcolor {gray!15 }
\raggedright Others** & 0.41&0.21\\
\midrule
\midrule
\end{tabular}
\begin{tablenotes}[para,flushleft]
 \item{Note: All values are in percentage. Chi - square test with Holm correction shows statistically significant difference (df=1, p $<$ 0.05) between male and female for all pieces of information except proof of place of work. * shows  p - value $<$ 0.01(df=1), ** shows p value $<$  0.0001}
  \end{tablenotes}
\vspace{-4mm}
\label{table:ans4}
\end{table}

We checked if different genders showed different concern about data collected and handled by service providers. Chi - square test with Holm correction showed that men and women had a different preference for sharing information with mobile service provider (see Table~\ref{table:ans4}). We found statistically significant difference (df=1, p $<$ 0.05) between men and women for sharing all kind of information except proof of place of work and contact number. Although different genders showed different preferences for sharing personal information with service providers, we found that both genders consistently showed higher confidence in mobile service providers than websites (see Table~\ref{tab:pi2013:govcop}) for data collected (Mann Whitney test,  p>0.05). 

\subsubsection{Improper access to personal information}
We asked participants if they agreed that \emph{consumers have lost all control over how personal information about them is circulated and used by the companies}. About 53\% participants agreed and 23.66\% strongly agreed with this statement. Participants felt that mobile service providers could allow improper access of personal information to third parties and government (see Table~\ref{tab:pi2013:govcop2}). When asked if mobile service providers can share their private information with government organizations without informing customers, 67.41\% participants agreed or strongly agreed with the statement. Opinion was no different for the government. About 55\% participants agreed or strongly agreed that government agencies could misuse information e.g. banking, phone records, property records, insurance, and income tax shared with them through UID project. This high percentage indicates that participants were worried about improper access by both government and mobile service providers.

\begin{table*}[!htbp]
\centering
\small
\caption{\small{Unauthorized / improper access of personal information to businesses, mobile service provider and Govt. There is an increased unrest among the citizens about information access than collection. $\mu$ represents mean rating on a likert scale of 5 and $\sigma$ represents the standard deviation. Rest of the values are in percentage. SA = Strongly Agree and A = Agree.}}
\setlength{\extrarowheight}{2pt}
\begin{tabular}{p{7cm} p{1.5cm} p{1.2cm} p{1.6cm} p{1cm}r p{1cm} p{1cm} p{.05cm}}
\midrule
\midrule
\centering
\small
\raggedright & \bf{Strongly agree/ Agree} & \bf{Neutral} & \bf{Strongly Disagree/ Disagree} & $\mu$&$\sigma$ & \bf{Male (SA+A)} & \bf{Female (SA+A)}\\
\midrule
\rowcolor {gray!15 } 
\raggedright Personal information and biometric data could be accessible to other private corporate through UID with whom you would NOT like to share otherwise (N=7,202) & 56.34&25.78 &17.87 &2.57&1.00&58.19&51.74\\
\raggedright Mobile service providers can share consumer private information with third parties (N=
10,305)  &47.31 &22.04 &30.66&2.81&1.23&47.12&47.40\\
\rowcolor {gray!15 } 
\raggedright Government agencies could have access to details e.g. banking, land records, and income tax records which can be misused by government agencies (through UID) (N=7,193) &55.92 &26.76 &17.95&2.53&0.98&55.52&54.70\\
\raggedright Phone conversations can be tapped by mobile service providers in national interest (N=10,363) &68.17&21.19 &10.62 &2.25&0.96&65.71&73.50\\
\rowcolor {gray!15 } 
\raggedright Mobile service providers can share the customer's information with government organization without informing the customers (N=10,366) &67.41 &19.90 &12.09&2.28&0.99&67.21&68.17\\
\midrule
\hline 
\end{tabular}
\vspace{-4mm}
\label{tab:pi2013:govcop2}
\end{table*}

We did not find statistically significant difference between the genders for concern on improper access to their personal information by governments and mobile service providers. On a likert scale of strongly agree to strongly disagree (5 levels), most participants -- female (48.05\%) and male (47.32\%) agreed that  phone conversations can be tapped by mobile service providers in national interest ($\chi^2$ = 2.8585, df = 4, p-value = 0.5818). Similarly we found no statistical difference between men and women on \emph{Government agencies accessing details e.g. banking, land records, and income tax records which can be misused by government agencies through UID} ($\chi^2$ = 2.4378, df = 4, p-value = 0.6558)

We found statistically significant different between participants' opinion on 1) Concerns for personal data collected by service providers such mobile service providers and government and 2) Improper access i.e disclosure of information by service providers to unauthorized parties (Wilcoxon Signed, p<0.001). Contrary to high trust in mobile service provider for protecting the private data they collected (see Table~\ref{tab:pi2013:govcop}), participants were concerned that mobile service providers may allow unauthorized access to personal information (see Table~\ref{tab:pi2013:govcop2}). Showing concern, 67.41\% participants agreed and strongly agreed that mobile service providers can share customer's information with government without informing them. However, showing confidence in the protection provided to data collected, 63.21\% participants agreed or strongly agreed that mobile service providers give reasonable protection to data collected. 

\subsection{Privacy and social sharing}
We asked participants \emph{with whom (friends, family, relatives, and society) would you share the following information} [a list]. Most participants were willing to share personal information with family members and were not willing to share any PII with society (see Table~ \ref{tab:pi2013:familyprivacy}).
Few participants (13.38\%) were willing to share information such as passwords (important PII) with family. Table~ \ref{tab:pi2013:familyprivacy} shows the difference between four groups --  friends, family, relatives, and society on sharing PII. Sharing preferences were statistically significant for each piece of PII (Cochran Test, p<0.0001). We further performed Post hoc Mc-nemar test and found statistically significant difference between various groups for sharing PII (except for landline numbers). There was no statistically significant difference for sharing landline numbers with friends or relatives. 

Hofstede found that a collectivistic society such as India shows high cohesion within groups and such societies share information with a larger group beyond family members and close friends~\cite{hofstedecluturalbook:1991}. 
However, in our survey, we found that participants shared personal information mostly with family members and few participants felt comfortable sharing personal information with friends and relatives (See Table~\ref{tab:pi2013:familyprivacy}). For example, 47.82\%  participants said they would share bank account details with family but 7.53\% and 1.10\% said that they would share bank account details with friends and society respectively.
Further, we found that participants' decision to share PII with family or society was dependent on how important they considered a piece of information. For financial information and passwords, we found a strong correlation (r$=$-0.95) between how important that PII was for a participant, and participant's willingness to share that PII with family. 
We found statistically significant difference between men and women for sharing information with friends, family, and society (Chi - squared test with Holm correction, df = 1, p<0.001). These percentages and difference in opinion shows that participants may share information with family but felt uncomfortable sharing information with friends and society at large.

\begin{table}[!htbp]
\small
\vspace{-4mm}
\caption{\small{PII sharing behavior with family, friends, relatives and society. All values are in percentages. }} 
\setlength{\extrarowheight}{2pt}
\begin{tabular}{p{2.1cm}p{1cm}p{1 cm}p{1.2cm}p{0.9cm}}
\midrule
\midrule
          & \bf{Friends} & \bf{Family} & \bf{Relatives} & \bf{Society}\\
  \midrule
   \raggedright Annual house hold income & 17.30  & 58.32 & 15.42 & 2.17\\
   \rowcolor {gray!15 } 
   \raggedright Bank account details** & 7.53  & 47.82 & 6.83  & 1.10\\
   \raggedright Credit card number & 5.02  & 38.65 & 3.75  & 0.91\\
    \rowcolor {gray!15 } 
  \raggedright  Date of birth & 31.38 & 40.03 & 27.91 & 7.59\\
   \raggedright Email address & 32.36 & 39.69 & 26.74 & 6.44\\
     \rowcolor {gray!15 } 
   \raggedright Family details & 35.47 & 50.46 & 39.22 & 6.47\\
   \raggedright Full name & 20.74 & 26.45 & 17.93 & 5.92\\
   \rowcolor {gray!15 } 
    \raggedright Health and medical history*** & 26.79 & 59.66 & 27.96 & 2.85\\
  \raggedright  Landline number* & 36.45 & 44.41 & 36.57 & 5.63\\
   \rowcolor {gray!15 } 
  \raggedright  Marital status & 27.06 & 34.05 & 25.14 & 7.84\\
  \raggedright  Mobile number & 39.54 & 43.80  & 34.90  & 5.57\\
    \rowcolor {gray!15 } 
  \raggedright  Passport number & 9.00 & 34.49 & 6.85  & 1.35\\
  \raggedright  Passwords & 2.10   & 13.38 & 1.30   & 0.52\\
  \rowcolor {gray!15 } 
   \raggedright Personal income & 12.10  & 41.99 & 9.43  & 1.43\\
   \raggedright Pictures and videos  & 44.81 & 56.53 & 34.74 & 3.02\\
   \rowcolor {gray!15 } 
   \raggedright Physical details & 33.97 & 50.75 & 29.84 & 3.74\\
   \raggedright Postal mailing address & 32.05 & 41.29 & 29.46 & 6.00\\
   \rowcolor {gray!15 } 
  \raggedright  Religion & 18.12 & 25.84 & 19.16 & 6.40\\
   \midrule
   \midrule
    \end{tabular}%
    \begin{tablenotes}[para,flushleft]
 \item{Note: Statistically significant difference exists between all four groups (Cochran's Q test, p$<$0.0001). Post hoc McNemar's test (after Bonferroni's correction). *, **, ***, show p=0.76, p=0.003, p $<$ 0.001 respectively when comparing friends and relatives. For all other combinations McNemar's showed significant difference among groups (McNemar's, p $<$ 0.0001).}
  \end{tablenotes}
    \vspace{-4mm}
  \label{tab:pi2013:familyprivacy}%
\end{table}

\vspace{-1mm}
\subsection{ Privacy attitudes influence technology use}
In this section, 1) we show privacy concerns can influence participants' willingness to use various services through mobile phone and OSN and 2) we find that users understand little about their actions that can result in privacy threats and thus show low preparedness to protect themselves.
\vspace{-1.5mm}
\subsubsection{Willingness to use technology}
\begin{figure}[!h]
 \vspace{-6mm}
\begin{center}
\includegraphics*[viewport= 2 2 470 300 , scale=0.45 ]{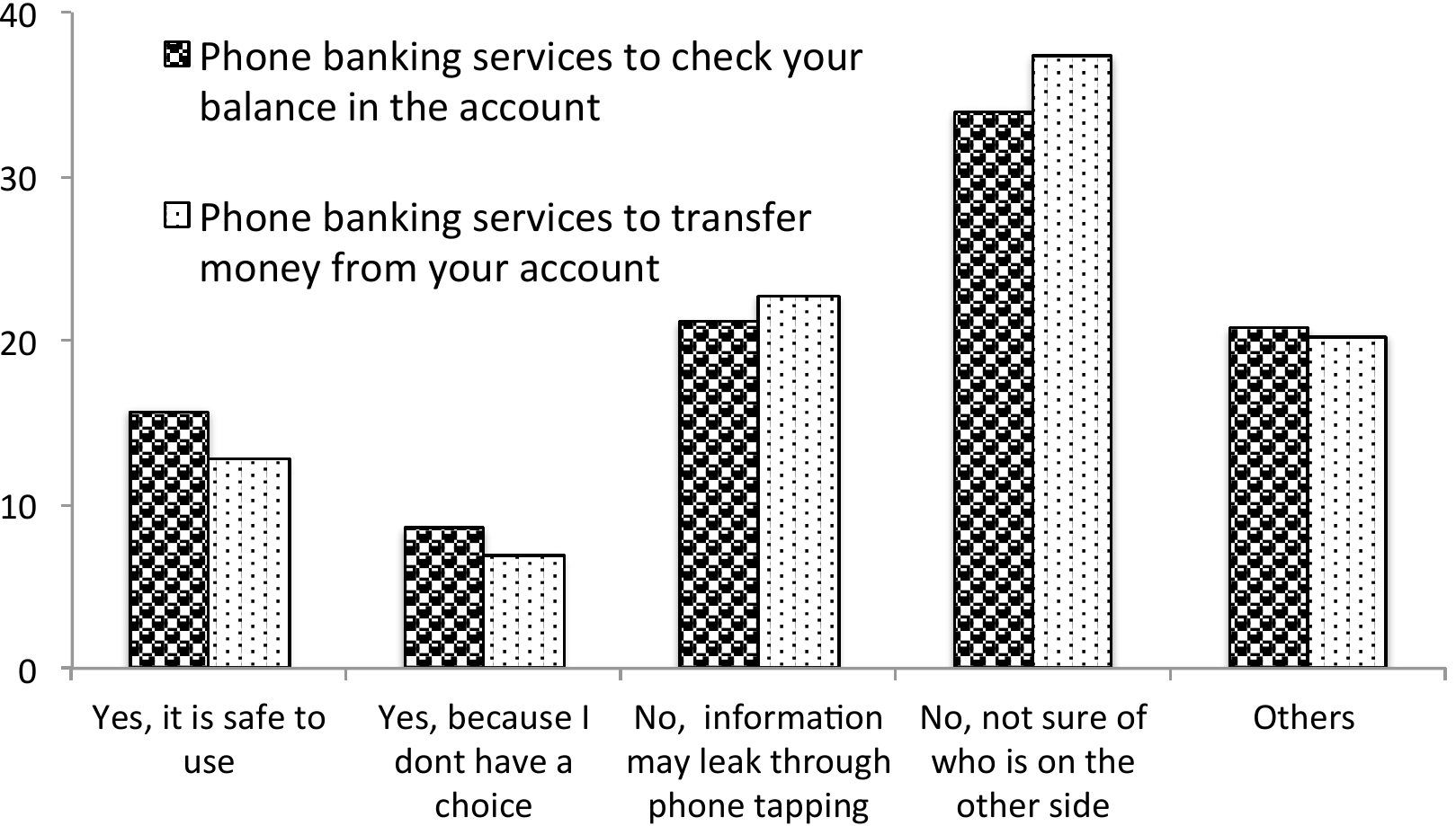}
\vspace{-3mm}
\caption{\small{Participants expressed concerns that information can be leaked through phone tapping; 33.93\% and refused to use phone banking as they were not sure who was on the other side of the call.}}
\label{fig:FPB}
\end{center}
\vspace{-6mm}
\end{figure}

We asked participants if they would use phone banking service to check their balance or to transfer money from their account. Among 10,349 \footnote{Some participants chose not to answer this question, therefore, the total number of responses were not 10,427.} participants, only 15.73\% said \emph{phone banking is safe to use to check balance}. However, 21.11\% expressed concern about ~information~being ~leaked through phone tapping and  33.93\% participants refused to use phone banking as they were not sure who was on the other side of the call. Participants expressed higher anxiety for transferring money using phone banking than checking balance. Only 12.77\% (N= 10,291) found phone banking to be safe for transferring money whereas 22.71\% mentioned concerns about fear of information being leaked through phone tapping. Moreover, 37.34\% mentioned concerns about not being sure of who was on the other side (nearly 3\% higher to the percentage of participants who mentioned same concerns for checking balance). Figure~\ref{fig:FPB} illustrates participants' concern and fear for using phone banking facilities. These observations show high anxiety among participants that may result from privacy concerns while using mobile phones.

We found a significant difference between men and women on using phone banking services to check the balance in their bank account ($\chi^2$ = 596.46, df = 4, p-value < 0.001). Higher number of women (41.73\%) than men (30.40\%) said that will not like to use phone banking for checking account balance as they were not sure of who was on the other side of the phone call. Table 9 shows male and female participants' concern for not using phone banking services to check account balance. Similar to perceptions for checking account balance, there was statistically significant difference ($\chi^2$ =  589.34, df = 4, p-value < 0.001) between men and women for using phone banking services to transfer money from their account (See Table 9). Higher number of women (46.65\%) were not sure of who was on the other side of the phone call while using phone banking services to transfer money than men (30.40\%). Aforementioned behavior indicates that participants were concerned about giving details to an unknown person, which could lead to a privacy breach. Thus, participants refrained from using technology that did not offer adequate privacy.

\begin{table}[!htbp]
\vspace{-3mm}
{\caption{ \small{Gender preference on the use of phone banking services to check the balance and to transfer money from one account to another. }}}
\small

\begin{tabular}{p{5cm}p{1cm}p{1cm}}
\midrule
\midrule
\bf{N=10,349}& \bf{Female}& \bf{Male}\\
\midrule
\multicolumn{3} {c} {\bf{Phone banking to check account balance}}\\
\midrule
 \rowcolor {gray!15 } 
Yes, it is safe to use & 13.01& 17.06\\
Yes, I don't have a choice & 11.34& 7.13\\
 \rowcolor {gray!15 } 
No, I fear information may be leaked through phone tapping & 25.99& 18.6\\
No, not sure of who is on the other side & 41.73& 30.4\\
 \rowcolor {gray!15 } 
Others & 7.92& 26.81\\
\midrule				
\multicolumn{3} {c} {\bf{Phone banking services to transfer money}}\\
\midrule
 \rowcolor {gray!15 } 				
Yes, it is safe to use & 9.76& 14.13\\
Yes, I don't have a choice & 9.43& 5.63\\
 \rowcolor {gray!15 } 
No, fear information may be leaked through phone tapping & 27& 20.56\\
No, not sure of who is on the other side & 46.65& 33.15\\
 \rowcolor {gray!15 } 
Others & 7.16& 26.53\\
\midrule
\hline
\end{tabular}
\label{tab:bankingcheck1}
    \begin{tablenotes}[para,flushleft]
 \item{Note: Statistical difference exists between genders to check balance ($\chi^2$ = 596.46, df = 4, p-value < 0.001) and transfer money ($\chi^2$=  589.34, df = 4, p-value < 0.001).}
  \end{tablenotes}
\end{table} 
 
\subsubsection{Low Preparedness to protect privacy}
Next, we studied how aware were participants of the actions that may result in a privacy breach. For this, we analyzed participants' preference for accepting friend requests on their most used OSN. This knowledge of participants' preference helped in judging their capabilities to make friends and also to keep themselves protected. Table~\ref{tab:acceptfriendrequest} shows participants' decision to accept friendship request from different people. About 28\% participants said that they would accept friendship request from unknown people who were of the same hometown, and 19.51\% would accept friendship requests from a person of the opposite gender. Moreover, 8.31\% would accept friendship requests from somebody whom they did not recognize but have mutual friends with. These preferences show that a considerable number of participants may accept friend request from random and unknown people. Such an attitude may interfere with information protection on OSN and indicate low preparedness level to handle privacy breach. These numbers show women were more conscious of the actions that may result in a privacy breach than men.

\begin{table}[!htbp]
\caption{ \small{Participants' response for accepting friendship request on their most frequently used OSN. All values are in percentage.}}
\centering
\small
\setlength{\extrarowheight}{2pt}
\begin{tabular}{p{5cm}p{2cm}}
\hline
\midrule
& N=6,929\\
\hline
Colleagues &61.97\\
 \rowcolor {gray!15 } 
Family Members&71.21\\
Friends&79.03\\
 \rowcolor {gray!15 } 
People from my hometown&27.39\\
Person of opposite gender&19.51\\
 \rowcolor {gray!15 } 
Person with nice profile picture&10.12\\
Strangers (people you do not know) &4.99\\
 \rowcolor {gray!15 } 
Somebody, whom you do not know or recognize but have mutual / common friends with &8.31\\
Anyone&2.99\\
 \rowcolor {gray!15 } 
Others&0.74\\
\midrule
\hline
\end{tabular}
\vspace{-1mm}
\label{tab:acceptfriendrequest}
\end{table}

\begin{figure}[!h]
\begin{center}
\includegraphics*[viewport= 2 30 420 260, scale=0.50]{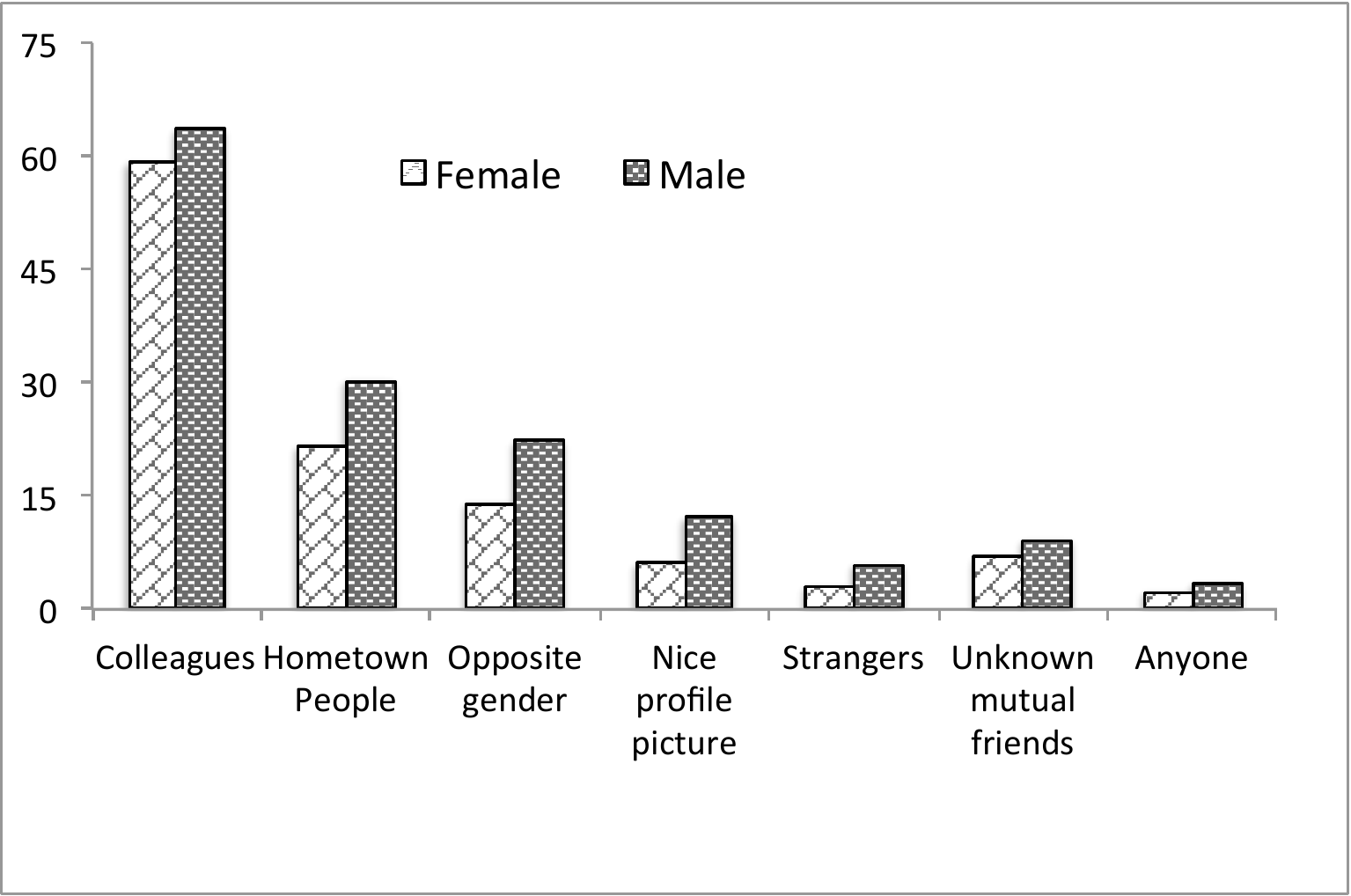}
\vspace{-4mm}
\caption{\small{Gender difference in accepting friend requests on OSN. Females were more conscious than males for accepting friend requests.}}
\label{fig:FM3}
\end{center}
\vspace{-6mm}
\end{figure}

Marginal difference exists in both the genders for accepting friend requests on OSN. We asked participants \emph{if they receive a friendship request on their most frequently used network, whom would they add as friends.} Participants chose their answer from the following options --  colleagues, family members, friends, people from hometown, opposite gender, nice profile picture, strangers, mutual friends, anyone or others. Figure~\ref{fig:FM3} illustrates male and female participants' preference for accepting friendship request from different individuals. In comparison to women, more male participants said that they felt comfortable while accepting friendship requests from colleagues, people from hometown, opposite gender, mutual friends and others (Chi squared test, p-value $<$ 0.001). In comparison to 14.10\% women, 22.20\% men (nearly 8\% more) said that they would accept friend request from a person of the opposite gender. About double the number of men than women said that they would accept friendship requests from people with nice profile picture. 
However, men were less likely to make friends with family members than women on OSN. 

%



\section{Discussion}
ICTD research shows that the rapid adoption of technology has resulted in privacy issues in developing countries such as India~\cite{Ben-David:2011:CSD:1999927.1999939, Dodson:2013:MGC:2516604.2516626, Kolko:2007:CIC:1242572.1242689, Srinivasan:2013:CMR:2516604.2516625}. However, these research works provide little understanding on \emph{what are end users' privacy expectations} and \emph{how users are concerned about their privacy} in India. In this work, we explore end-users' privacy perceptions and expectation to understand their privacy needs while using technology such as mobile phones and OSN. To this end, we conducted surveys with 10,427 Indian respondents spread all around the country. 

Though preliminary, our results show that Indian participants disproportionately find financial details and passwords as important PII than health records or physical details. We find that participants do not feel comfortable sharing PII such as financial details with society. Participants do not feel comfortable sharing personal information such as passwords even with family members. Another recent study has observed similar concerns among Indians using OSN~\cite{reference8}. Though ICTD research shows end-users in the developing world share devices such as mobile phones within communities, we find that this may not be true for the information stored on the mobile phones. Participants in our study do not feel comfortable sharing information (especially financial details and password) with society and friends. The ICTD researchers work to design medical and financial technologies such as m-finance\footnote{improved mobile support for financial or banking facilities} to improve life in developing countries. Researchers creating these technologies or crowd-sourced ICTs or appointing field workers to collect critical PII  may need to consider these user preferences for increasing end - users' trust and acceptance of these technologies.

Meeting privacy expectations of the end-users can help to improve technology acceptance and thus improve life in developing nations. In our study, participants say that they refrain from using technology such as phone banking services as \emph{they are not sure of whom they are talking to} and this unsurety may result in a privacy breach. Despite these privacy concerns, we find that participants are not much prepared to protect themselves from a privacy breach. For instance, while users were not willing to share information with anyone in the society, they accepted random friend requests from unknown people who may be from their home town  or of the opposite gender. Thus, there is a mismatch between users' expectations of privacy and the actual actions that guarantee privacy. Therefore to protect users, regulatory bodies and technologies should try to encounter this mismatch by educating users on the risks involved. 

Similar to existing studies~\cite{bellmaninternationaldifferences:2004, ecommerce1998}, our results also show that different genders have different privacy needs and expectations. Women in our study are marginally more conscious of privacy than men.  Differences in perceptions of privacy among genders show that the technology, storage policy and system level designers cannot expect one-size-fits-all solution that can accommodate gender preferences. In this study, although we present difference in gender preferences, such differences may exists in other social dynamics like income, education and other factors such as age. We aim to study these aspects in future work.

We find that participants think that government may misuse personal data collected using ICTs during governance projects such as UID. Participants also worry that mobile service providers may share private information with government without their consent.  Clearly, there is a concern among users for privacy provided by government services using ICTs.  These ~concerns ~may ~be ~challenging for success of ~government ~~projects that plan to reap the benefit of use of technology such as mobile phones and OSN. In India, the government is increasingly introducing e-governances facilities (using mobile phones and OSN) for citizens. These services include UID project, lost and found case reporting, traffic police OSN pages, city police OSN pages, and the municipal corporation's administrative services. These services rely on the fact that users will share data on online and open platforms. To foster development and wider service adoption, regulatory bodies and technologists alike should try to reduce this fear in users by meeting users' expectation. If privacy concerns do not impede adoption of such technologies, these technologies can provide better security and standard of living in developing nations.

We hope that the findings in our paper will be relevant to researchers, practitioners and policymakers who are involved in improving the use of privacy oriented technologies for the benefit of people in developing region. Also, these findings may be relevant to researchers designing privacy-preserving technologies in other developing world communities similar to India. Furthermore, though this is not a policy paper, we hope that technologists in the privacy policy domain may get useful insights for devising procedures that may help to address the growing problem of privacy while using ICTs.

\section{Acknowledgments}
We would like to thank International Development Research Centre (IDRC), Canada for supporting the research through a Pan-South-Asia project on privacy. We thank all participants for their efforts and sharing their thoughts with us, during the study. We would like to thank all the members of Precog and CERC who have given us continued support throughout the project; special thanks to Aditi Gupta, Srishti Gupta, Siddhartha Asthana, Deepansha Sachdeva and Anuradha Gupta. Many others around the country have helped us with
data collection, we would like to express our thanks to everyone.

\bibliographystyle{abbrv}
\small
\bibliography{u2p2} 
\end{document}